\documentclass[prb,twocolumn,showpacs,preprintnumbers,amsmath,amssymb]{revtex4-1}

\usepackage{graphicx}
\usepackage{dcolumn}
\usepackage{bm}

\begin{document}

\title{Electronic Structure and Upper Critical
Field of Superconducting Ta$_2$PdS$_5$}

\author{David J. Singh}

\affiliation{Materials Science and Technology Division,
Oak Ridge National Laboratory, Oak Ridge, Tennessee 37831-6056}

\date{\today}

\begin{abstract}
We report electronic structure calculations for Ta$_2$PdS$_5$, which is a
layered superconductor containing heavy elements and displaying an
upper critical field, $H_{c2}(0)$,
$\sim$3 times higher than the estimated Pauli 
limit.
We show that this is a multiband superconductor that is
most likely in the strong coupling regime.
This provides an alternative explanation to strong spin orbit
scattering for the high upper critical field.
\end{abstract}


\maketitle

\section{introduction}

Lu and co-workers
\cite{lu}
recently reported fully gapped type-II superconductivity with
$T_c$ up to 6 K in the layered chalcogenide Ta$_2$Pd$_x$S$_5$, $x\alt$1.0.
Remarkably, they found anisotropic upper critical fields,
$H_{c2}$, up to 31 T.
This is substantially above the estimated Pauli limit of $\sim$10 T.
They argued that the high critical fields are the result of
strong spin-orbit scattering associated with Pd vacancies.

The purpose of this paper is to present the electronic structure
in relation to the experimental findings and in particular the
high upper critical field.
We find that the compound is
a three dimensional low carrier density metal, with moderate anisotropy and
a Fermi surface structure that can support multiband superconductivity.
This is qualitatively similar to the findings of recent studies
of the niobate superconductors,
Nb$_2$Pd$_{0.81}$S$_5$ and Nb$_3$Pd$_{0.7}$Se$_7$,
which also show high upper critical fields. \cite{zhang,zhang2}
A comparison of the calculated density of states with the experimental
suggests this compound is in the strong coupling regime.
Multiband, strong coupling superconductivity provides and alternative
to the proposed spin orbit scattering explanation of the observed
high $H_{c2}$ values.

\section{approach and structure}

The present calculations were done for the stoichiometric compound.
They were based on density functional theory,
using the generalized gradient approximation (GGA) of Perdew, Burke and
Ernzerhof (PBE). \cite{pbe}
For this purpose we used the general potential linearized augmented
planewave (LAPW) method, \cite{singh-book}
as implemented in the WIEN2k code. \cite{wien2k}
We used well converged basis sets including local orbitals to treat
the semicore states with LAPW sphere radii of 2.25 bohr and 2.1 bohr, 
for the metal and S atoms, respectively. We used the lattice parameters
measured by Lu and co-workers, \cite{lu} and relaxed the atomic positions
in the cell by total energy minimization.
This relaxation was done
with relativity included at the scalar level. Once the atomic
positions were obtained, the electronic structure was calculated
relativistically, including spin-orbit.

\begin{figure}
\includegraphics[width=0.9\columnwidth,angle=0]{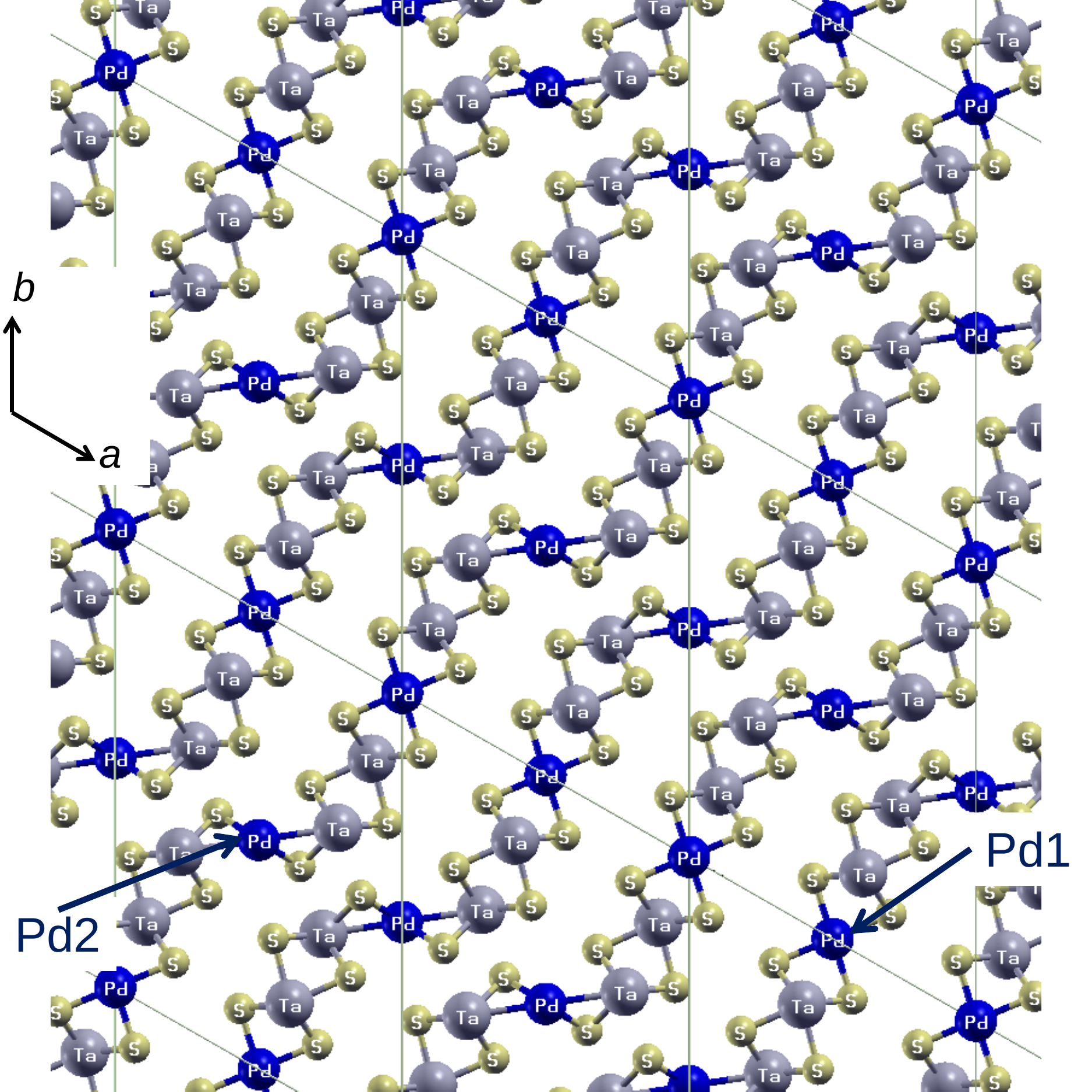}
\caption{(color online) Structure of Ta$_2$PdS$_5$ showing the two
different Pd sites. The atomic positions as as obtained from the
relaxation. The $a$ and $b$ directions for the conventional cell
of the centered monoclinic cell are as shown; the $c$-axis is perpendicular
to the plane depicted. The lines indicate the conventional unit cell.
}
\label{struct}
\end{figure}

Ta$_2$PdS$_5$ is made up of corrugated metal sulfide sheets as
shown in Fig. \ref{struct}.
The crystal structure is centrosymmetric, space group No. 12 ($C2/m$),
two formula units per primitive cell,
and contains two different Pd sites, Pd1 and Pd2 as well as two
different Ta (Ta1 and Ta2) and five different S sites.
All the metal sites form chains along the $c$-axis direction, with spacing
of $c$=3.269 \AA. In the following we use the standard $C2/m$ setting,
which differs from the setting used in Ref. \onlinecite{lu}. In particular,
the high conductivity direction, labeled as the $b$-axis, in 
Ref. \onlinecite{lu}, is the $c$-axis direction with the setting used here.

The structure
may be described starting with the Pd1 atoms.
These are nearly square planar coordinated by S.
These near squares are stacked along the $c$-axis direction, with a 
spacing of $c$=3.269 \AA, and edges perpendicular and parallel to the
flat part of the sheets shown in Fig. \ref{struct} (i.e. S above
and below the sheet).

The Ta1 and Ta2 atoms are also in chains along
the $c$-axis direction, with the Ta1
chains adjacent to the Pd1 and the Ta2 at
the edge of the flat parts of the sheets.
Along the $c$-axis the Ta1 are between the Pd1S$_4$ squares, and each of the
S in these squares then bonds to two Ta1 atoms and one Pd1.
The Ta1 site is six fold coordinated by S. Four of these are from
the Pd1S$_4$ squares, and two are shared with the Ta2 sites,
completing the flat parts of the sheets. The Pd2 site is
on the step part of the corrugated sheets and is also in an approximately
square planar coordinated site, but in the S coordinating the Pd2
are shared between Pd2 atoms along the $c$-axis chains. Each of these
S is shared between two Pd2 atoms, and is also bonded to two of the Ta2
atoms. This leads to longer S-Pd2 bond lengths of $\sim$2.42 \AA, as
compared to the Pd1-S distances of $\sim$ 2.33 \AA.
These edges also connect the sheets because of relatively short S-S
distances.
Experimentally, the Pd2 site has a variable number
of random vacancies.

\section{results and discussion}

\begin{figure}
\includegraphics[height=0.9\columnwidth,angle=0]{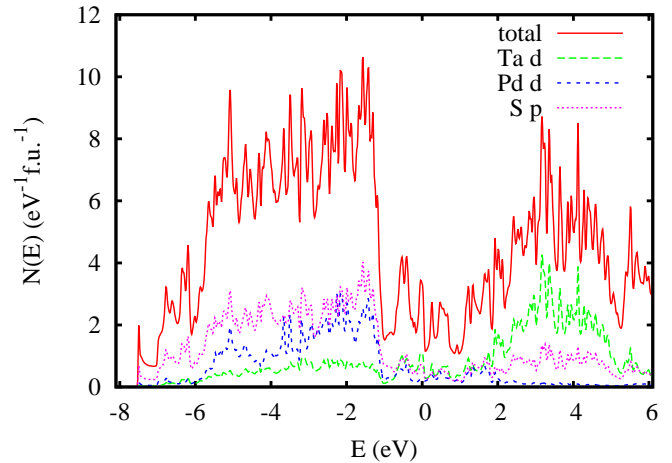}
\includegraphics[height=0.9\columnwidth,angle=0]{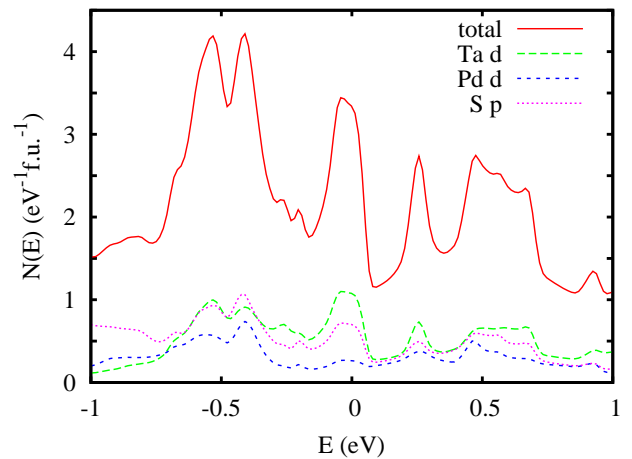}
\includegraphics[height=0.9\columnwidth,angle=0]{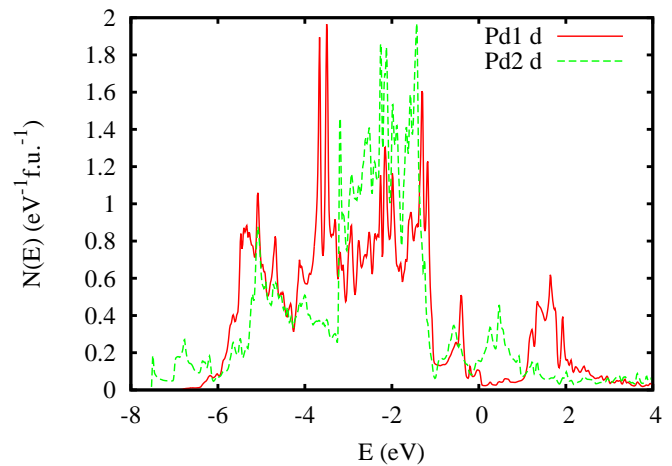}
\caption{(color online) Electronic density of states and transition
metal $d$ and S $p$
projections (top), blow-up around $E_F$ (middle)
and $d$ projections onto to the two different
Pd sites (bottom).
Note that the plots are on a per formula unit basis
and that there are two formula units per primitive cell, as
1/2 Pd1 and 1/2 Pd2 atoms per formula unit. The energy zero
is at $E_F$.  }
\label{dos}
\end{figure}

We begin by presenting the electronic density of states (DOS).
The calculated
electronic density of states and projections onto the various sites
are shown in Fig. \ref{dos}.
As may be seen, there is strong covalency between the Ta $d$ and various
S $p$ states. 
The Pd $d$ contribution to the electronic structure is mostly in the
occupied bands.
However, each Pd contributes one band that is split off above the
main Pd $d$ density of states. This band has Pd $d$ - S $p$ $\sigma$
antibonding character and is mirrored by corresponding bonding states
at the bottom of the Pd DOS (note that
this bonding character
is spread between several hybridized
bands). In any case, this structure, with the
antibonding band above the Fermi level, reflects the bonding of the Pd to
the S comprising the square planes.

The Pd1 - S $\sigma$ antibonding band mentioned above extends from
$\sim$ 1 eV to $\sim$ 1.9 eV above the Fermi energy, $E_F$.
In contrast for the Pd2 site, the $\sigma$ antibonding band is
broader and partially occupied extending from $\sim$ -1 eV to $\sim$ 1 eV,
with respect to $E_F$. This partial occupation of the antibonding band
means that the Pd2-S bonding is weaker than the Pd1-S bonding, which
may explain why the Pd2 site is susceptible to vacancies.
It also means that while the electronic structure near $E_F$,
relevant to transport and superconductivity,
has mainly hybridized Ta $d$ - S $p$ character,
it also includes a significant
Pd2 $d$ - S $p$ antibonding contribution.
In contrast, there is very little Pd1 character near $E_F$.

\begin{figure}
\includegraphics[width=\columnwidth,angle=0]{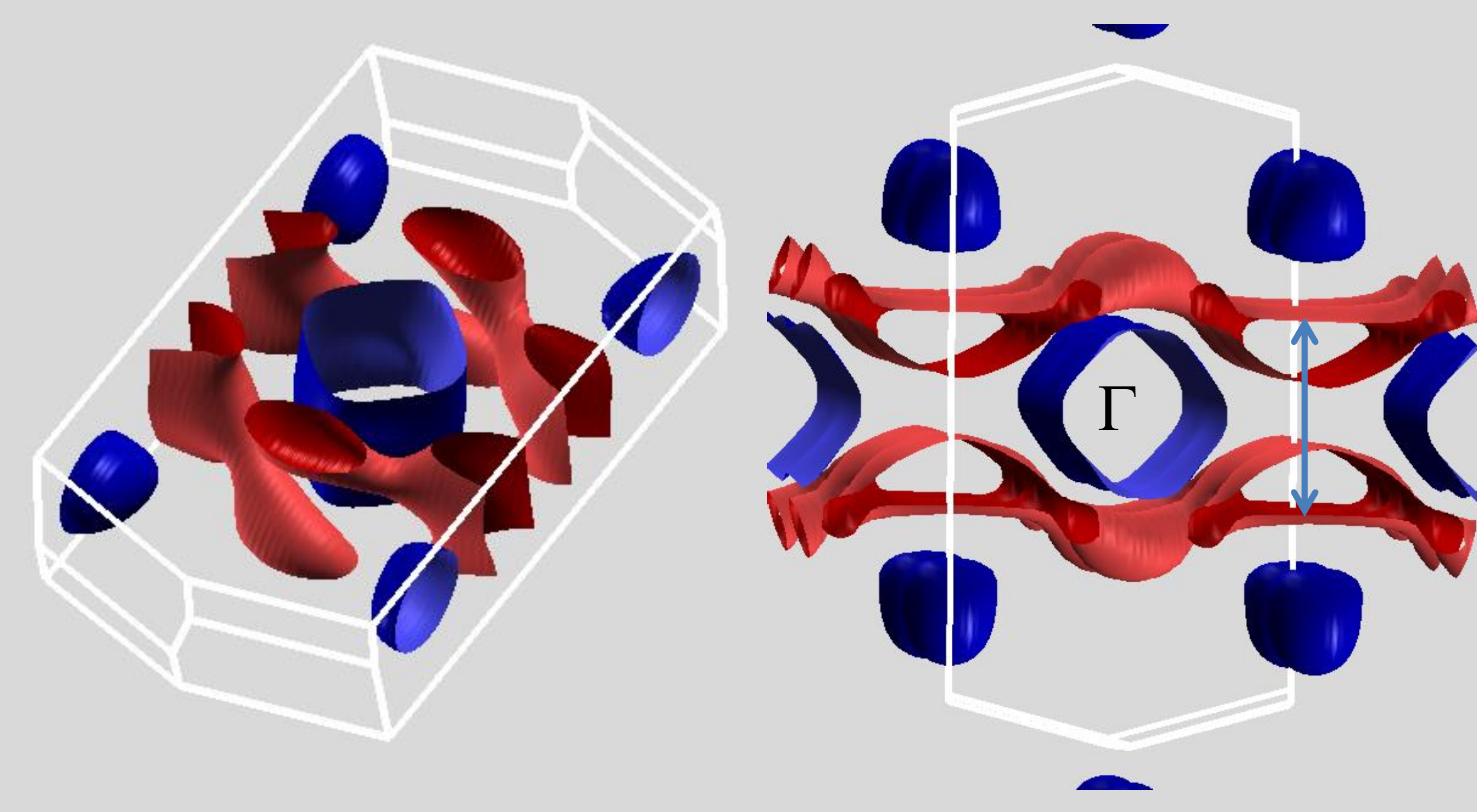}
\caption{(color online) Calculated Fermi surfaces of Ta$_2$PdS$_5$
including spin-orbit.
Electron-like sections are shown as blue and hole-like sections are shown
as red.
The right panel shows a view along the
$k_y$
($b^*$) direction with an extended zone to show the structure of the zone
boundary electron surface, while the left panel is at an oblique angle.
The white lines denote the primitive Brillouin zone. The $\Gamma$ point
is drawn at the center.
The arrow on the right panel indicates
the nesting vector for the flat parts of the hole sections.}
\label{fermi}
\end{figure}

The calculated Fermi surface is shown in Fig. \ref{fermi}. It consists
of three sheets. The first is a hole sheet that consists of
flattened cylinders running along zone boundary in the $k_y$ direction,
with complex interconnection across the zone. The outer parts of the 
cylinder are flat indicating a nesting at the vector that connects the
these, as shown in the figure
(but note that the strong defect scattering related
to Pd2 vacancies in this material
may obscure this nesting).

This hole sheet is compensated by two
electron sheets. These are a very two dimensional cylinder around
the zone center and small closed sections on the zone faces as shown.
This characteristic of compensating hole and electron Fermi surfaces
is somewhat reminiscent of the Fe-based superconductors.
\cite{singh-fesc}
However, there is good experimental evidence that the physics here is very
different, and specifically that Ta$_2$Pd$_x$S$_5$ is an $s$-wave electron
phonon superconductor. \cite{lu}

\begin{figure}
\includegraphics[width=\columnwidth,angle=0]{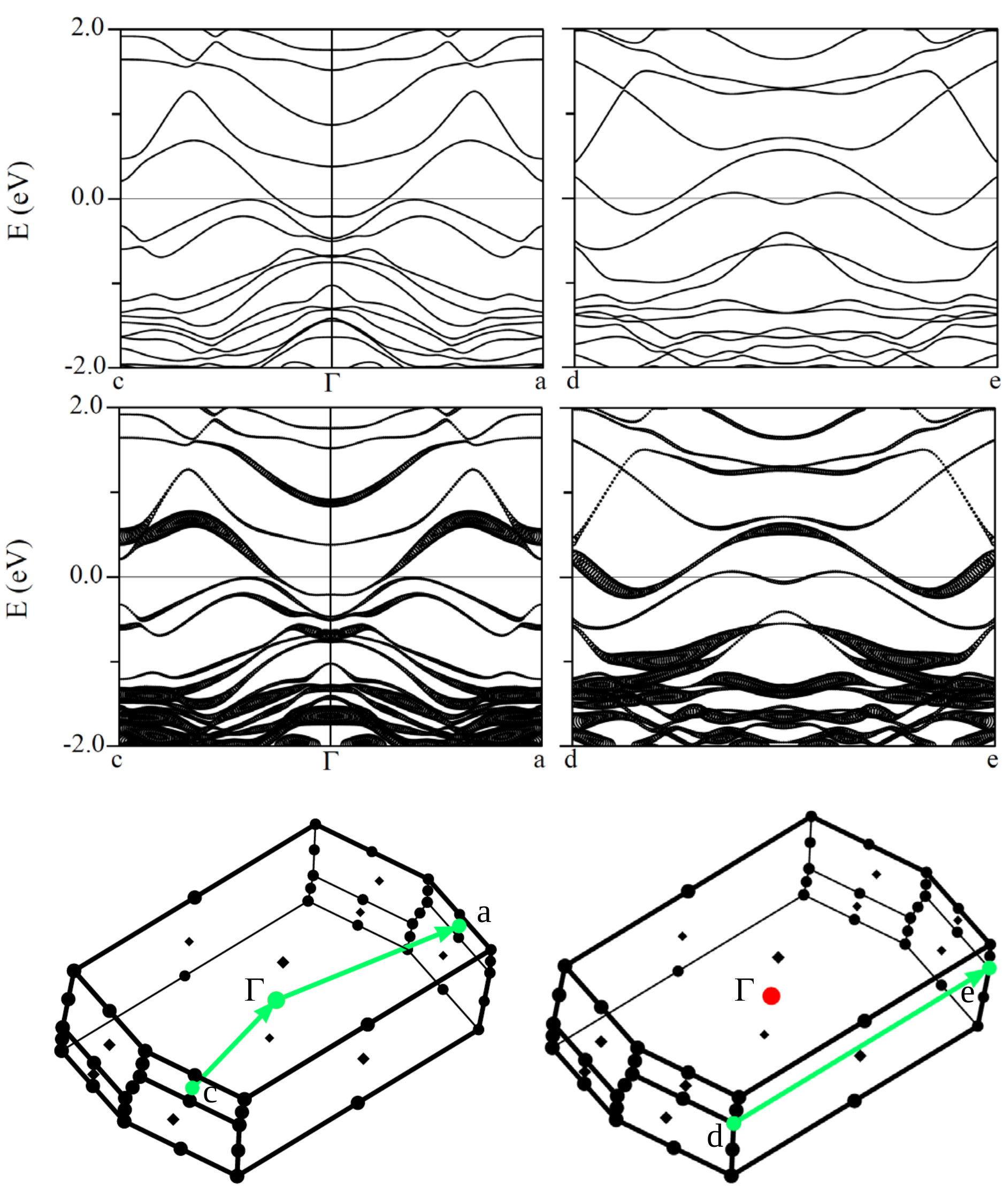}
\caption{(color online) Band structure plotted
with lines (top) and dots weighted by the Pd2 $d$ character (middle),
along different lines as shown (bottom).
The bands are plotted along non-symmetry lines in
order to show the crossings of the various Fermi surfaces.
The energy zero is at $E_F$.
Note that the large face of the primitive zone is perpendicular to
$b^*$, while the points a and c are along the $a^*$ and $c^*$
directions, respectively.}
\label{bands}
\end{figure}

The calculated DOS at the Fermi energy, $N(E_F)$=3.34 eV$^{-1}$ on
a per formula unit basis.
According to the Sommerfeld formula,
$\gamma_{\rm bare}=(\pi^2k_B^2/3)N(E_F)$,
this corresponds to a bare
specific heat coefficient, $\gamma_{\rm bare}$=7.88 mJ/(mol K$^2$) again
per formula unit. Comparing with the experimentally reported
value,
$\gamma_{\rm expt.}$=27.6 mJ/(mol K$^{2}$)
one may estimate a specific heat enhancement,
(1+$\lambda$)=$\gamma_{\rm expt.}$/$\gamma_{\rm bare}$=3.5,
or $\lambda$$\simeq$2.5.
An electron phonon coupling in this range characterizes Ta$_2$Pd$_x$S$_5$ as a
strong coupling superconductor.
Strong coupling superconductors have a larger ratio of the gap to the
critical temperature, i.e. $2\Delta/kT_c$, above the weak coupling
(BCS) value and renormalization of the normal state, both of
which can increase the higher Pauli limit on the
upper critical field.
\cite{clogston,rainer,orlando}
Using the formula,
$H_P(0)/H_P^{\rm BCS}(0)=\eta_{H_c}(0)(1+\lambda)^{1/2}$,
from Ref. \onlinecite{orlando},
and setting the parameter $\eta_{H_c}(0)$, which is larger than unity,
to unity, one obtains a low estimate of the strong coupling Pauli limit,
$H_P(0)$ as $\sim$1.9 times the weak coupling BCS value $H_P^{\rm BCS}(0)$.
Thus, based on this
analysis, it is quite likely that strong coupling effects in Ta$_2$PdS$_5$
can readily yield a factor of two or more increase in the Pauli paramagnetic
limit, with no further considerations.

The conductivity anisotropy as obtained
in the constant scattering time 
approximation (CSTA) with the BoltzTraP code,
\cite{boltztrap}
shows that the in-plane anisotropy is weak.
The lowest conductivity direction is predicted to be close
to the $b$-axis, across the layers. 
If the scattering rate is independent of direction, as is typically
valid in metals, the average conductivity in the high
conductivity $a$-$c$ plane
will be 2.3 times larger than in the $b$-direction, and have little anisotropy
itself. The highest conductivity direction is predicted to be $c$, which
is the direction of the metal chains, but only
by $\sim$2\% compared to the highest conductivity direction
in the $a$-$b$ plane. This may be sensistive to disorder due to 
Pd2 vacancies, which would affect the $a$-$b$-plane more than the
$c$-axis direction.
In any case, the result is consistent with the inference of Lu and co-workers
(note that the $c$-axis here is the same as the $b$-axis in the
setting used by Lu and co-workers). \cite{lu}
Lu and co-workers found the highest $H_{c2}(T)$ along this direction
($c$-axis as defined here),
\cite{lu}
which is consistent with the fact that it is the high conductivity
direction.

However, the experimental $H_{c2}$ anisotropy of $\sim$2.3 is larger than
that which would be expected from a single band based on the conductivity
anisotropy, i.e. $(\sigma_{ac}/\sigma_b)^{1/2}\simeq1.5$.
In a very dirty material with very strong point defect scattering, the
assumption of the CSTA, that all bands have scattering with
the same scattering rate, $\tau^{-1}$ may not be valid.
On the one hand the band structure shows that the hole
band has a pudding mold shape (see especially the right
panel of Fig. \ref{bands}) that places $E_F$ very close to the
band edge. This suggests that this band and the small electron 
pockets on the zone faces could be easily Anderson localized
(this is controlled by the the strength of the disorder
potential relative to energy distance of the band edge
from the Fermi energy). \cite{tvr}
This would increase the relative contribution of the rather 2D
electron cylinder at the zone center to transport and thereby
increase the anisotropy.

On the other hand, if the strength of the scattering is not
enough to produce Anderson localization but the
point defect scattering is still very strong
in the strength of providing a mean free path
set by the defect spacing, it could be that
the mean free path, $l=v_F\tau$ for the different bands
may become similar. If that is the case in the samples of Ta$_2$Pd$_x$S$_5$
that were reported,
the lighter electron cylinder section of the Fermi surface
would be less dominant in the conductivity.
In any case, from a transport point of view,
Ta$_2$PdS$_5$ should be classed as a three dimensional metal, with
a moderate planar anisotropy.
It will be of interest to perform transport measurements to determine
the conductivity anisotropy in relation to the CSTA
results, if suitable superconducting samples can be prepared.

Returning to the Fermi surfaces, the band structure along directions
cutting the different sheets and the Pd2 character are shown in
Fig. \ref{bands}.
Both the electron and hole sections come from bands of hybridized Ta $d$
and S $p$ character. However, the band giving rise to the hole cylinder
around the zone center shows much stronger dispersion than the that
comprising the electron sections.
Of the three sections of Fermi surface, the small electron ellipsoids
on the zone faces have the highest Pd2 character. This combined
with there small size may make them particularly sensitive to Pd2
vacancy induced disorder.

In any case, superconductivity is likely to be dominated by the two larger
sheets.
While, as mentioned, there is little in plane anisotropy for the total
conductivity the individual band anisotropies are different. For the
$b$-axis (low conductivity direction) the electron bands
(the cylinder and the electron pockets on the zone faces together) contribute 
$\sim$65\% of the conductivity. For the $a$ axis direction this
is more than $\sim$85\%, and for the $c$ direction it is $\sim$55\%.
In any direction, the electron sheets are lighter in the sense of having
stronger dispersion and therefore greater conductivity than the hole sheets.
Thus while the hole sheets provide the larger contribution to $N(E_F)$,
the electron sheets provide the larger contribution to transport.

Ta$_2$PdS$_5$ should therefore be classed as a two band (ignoring the
small electron pockets) strong coupling superconductor, which is in the
dirty limit. In two band superconductors, one may expect an increase in
$H_{c2}(0)$ and a temperature range with a quasi-linear $T$ dependence.
The data of Lu and co-workers do in fact show a noticeable quasi-linear
dependence, especially for fields in the $c$-axis direction ($b$-axis
in their setting).
In such cases, MgB$_2$ a well studied examples,
\cite{gurevich2,gurevich}
defect scattering can lead to a substantial increase in $H_{c2}(0)$
without requiring spin orbit.
\cite{hunte,gurevich3,petrovic}

\section{summary and conclusions}

We present electronic structure calculations for Ta$_2$PdS$_5$
in relation to the observed superconductivity of Ta$_2$Pd$_x$S$_5$.
We find that this materials is a multiband superconductor and should
be classed as an anisotropic 3D metal. The conductivity anisotropy
is consistent with the observed high $H_{c2}$ direction.
A comparison of the density of states with the experimental specific
heat implies that this is a strong coupling superconductor.
The results indicate that the apparent violation of the single band
Pauli limit on $H_{c2}(0)$ found in this material may be due
to a combination of strong
coupling and multiband effects in the extreme dirty limit.
It will be of interest to determine to what extent these give rise to the
observed $H_{c2}$ and to what extent defect induced spin orbit scattering is 
important.
In any case, Ta$_2$Pd$_x$S$_5$ appears to be an excellent experimental
model for probing the interplay of multiband superconductivity, strong
coupling and strong spin-orbit scattering.

\acknowledgments

This work was supported by the U.S. Department of Energy,
Basic Energy Sciences, Materials Sciences and Engineering Division.

%

\end{document}